\documentclass[10pt,prd,aps,preprintnumbers]{revtex4}
\usepackage[dvips]{graphicx,color}
\begin{document}
\vspace*{.25in}
\title{ \LARGE{Strange Quark Nuggets in Space: Charges in Seven Settings}}         
\vspace*{.31in}
\large{
\author {E. S. Abers$^a$, A. K. Bhatia$^b$, D. A. Dicus$^c$, W. W. Repko$^d$, D. C. Rosenbaum$^e$, V. L. Teplitz$^{b,e}$}

\affiliation{\normalsize{$a$Department of Physics and Astronomy, UCLA, Los Angeles, CA 90095-1547 \\$^b$NASA Goddard Space Flight Center, Greenbelt, MD 20771 \\$^c$Physics Department, University of Texas, Austin, TX 78712
 \\$^d$Department of Physics and Astronomy, Michigan State University, East Lansing, MI 48824\\\\$^e$Physics Department, Southern Methodist University, Dallas, TX 75275} }         
\maketitle}
\normalsize{
\noindent{\bfseries Abstract.} We have computed the charge that develops on an SQN in space as a result of balance between the rates of ionization by ambient gammas and capture of ambient electrons.  We have also computed the times for achieving that equilibrium and binding energy of the least bound SQN electrons.  We have done this for seven different settings.  We sketch the calculations here and give their results in the Figure and Table II; details are in the  Physical Review D.79.023513 (2009).} 

\vspace*{.25in}
\begin{center}{\bfseries\large{ INTRODUCTION
}}\end{center}
\large{
It is now three dozen years since Bodmer's paper [1] suggesting that, while matter made of just up and down quarks is not stable against decay into baryons, matter made of u, d, and s might be.  It is two dozen since Witten's [2]
 seminal paper hypothesizing production of nuggets of strange quark matter (SQM) in the early universe when a phase transition might have caused strange quark nuggets (SQNs) to condense out of the quark soup.  But there has been no experimental or observational sign of SQM --- in spite of a reasonably vigorous program of searching. It is, however, important to pursue as wide a range of tests as possible, both for potential SQN discovery and for, lacking that, placing tighter bounds on loci in parameter space where SQM could be hiding.

The basic idea in concluding that quark matter with 3 kinds of quarks is more likely to be stable than quark matter with only two is that the potential energy of a system with $N_q$ quarks is the same independent of what flavor the quarks have.  It only depends on what colors they have.  However the kinetic energy of the system depends on both colors and light flavors (i.e., on how many fermi seas).  Thus, by the Pauli principle, for fixed quark total, the more kinds of flavors, the lower the fermi energies and the higher the likelihood of binding.  Introducing the c quark doesn't help because its mass is greater than the nucleon mass therefore costing more energy than is gained by adding another flavor.

Phenomenologically, the most important feature of SQM is its nuclear density, roughly $5\times10^{14}$g\,cm$^{-3}$.  Also, as Farhi and Jaffe showed [3], if SQM is, in fact, stable then, as the number of quarks in the system increases the binding energy per quark increases to an asymptotic value. Based on these properties deRujula and Glashow [4] sketched out what could be learned from a variety of observations and experiments if the dark matter (DM) in the galaxy (about $10^{-25}$g\,cm$^{-3}$) is SQM in the form of SQNs of mass M. Alcock, Farhi and Olinto [5] studied the properties of ``neutron stars'' if they are actually strange quark stars (SQSs).

A recent development of importance was the recognition that the attractive, flavor blind quark-quark color force leads to Cooper pairing [6] adding significantly to the binding energy of the system.  However, on the negative side, experiments and observations have lagged far behind theory.  They have failed to detect any evidence for SQM existence.  Some past, present and future SQM searches are summarized in Table I.  Beside the absence of SQN detection, there are two sets of observations that cast doubt on the prediction [5] that neutron stars are actually SQS's.  The first is the longstanding problem of pulsar glitches.  These are sudden changes in pulsar frequency, successfully understood on the basis of quakes in the crust of nuclear (not quark) matter around the neutron matter core.  They are not easily understood on the basis of an SQN core, for most models, which can only support a crust too thin for the quakes needed [5]; recently, however, theory has managed to produce a ``specialized'' SQN model which is capable of large enough quakes [7], but there does not yet seem to be any other confirmable predictions of the model.  Another problem with SQS's is that from  Fourier analysis of quasi periodic oscillations (QPOs) in``superbursts,'' nuclear explosions on soft gamma ray repeaters.  The frequencies have been shown [8] also  to indicate a crust depth greater than the limit found by [5].  

The object of the present work is to compute the charges that should develop on an SQN in space subject to ambient, ionizing radiation and deionizing capture of ambient free electrons. As a basic step in devising SQN space searches.

\begin{table}
\noindent
\caption{\normalsize{Some Strange Quark Nugget Searches. References for the data are found in (a)Sandweiss [9], (b)Price [1], (c) Spiering [11], (d) Banerdt \textit{et al.}[12], (e) Herrin \textit{et al.} [13]. }}

\begin{ruledtabular}
\noindent \begin{tabular}{lll}

   {\bfseries Experiment/Observation}& {\bfseries  Mass Range} (g) &{\bfseries Result} \\
\hline
AMS\footnotemark[1] &$10^{-24} - 10^{-22}$ & not done\\
RHIC\footnotemark[1]&$< 3\times 10^{-21}$ & not found\\
Mica Tracks\footnotemark[2]&$10^{-20} - 10^{-14}$ &$<<\rho_{DM}$ \\
ICE CUBE\footnotemark[3]  &$10^{-3} - 10^{-2}$&not done\\
Seismometers:&&\\
\,\,\,Future Lunar \footnotemark[4]  &$10^{3} - 10^{6}$&not done \\

\,\,\,Apollo\footnotemark[5]&$10^4 - 10^6 $& $< \rho_{DM}/10$\\
  \,\,\,USGS Reports\footnotemark[3]& $10^6 - 10^8$&$< \rho_{DM}$ \\
\end{tabular}
\end{ruledtabular}
 
\end{table}

\vspace*{.25in}
\begin{center}{\bfseries\large{ CALCULATIONS
}}\end{center}

We considered seven ``settings'' with both ionizing radiation producing net charge on the SQN and ambient electrons neutralizing that charge. The rate for the first falls with increasing $Z_N$ (number of electrons ionized; hence total positive charge of quark lattice with remaining electrons), while that for the second starts from near zero and rises with increasing $Z_N$.  We found the $Z_N$ for equality.  The seven settings are, in the order of decreasing equilibrium charge:  (1) the sun shining on an SQN near the Earth during a solar X-ray flare; (2) the (extra-galactic) diffuse background radiation (DBR) shining on an SQN in intergalactic space; (3) the diffuse background radiation shining on and SQN near the Earth; (4) an SQN in the primordial universe at recombination; (5) the galactic diffuse radiation (GDR) at the center of the galaxy (COG) shining on an SQN located near the COG; (6) the GDR shining on a solar system SQN near the Earth; and (7) an SQN in the CMB today (ignoring the DRB and other radiation).
Our model for the structure of the SQN is the so-called ``color-flavor locked'' model [6] in which there is Cooper pairing of quarks sufficiently strong that, in the bulk of the SQN the quarks occur in trios with zero charge, no color and antisymmetry in space wave functions.  All the net quark charge occurs on the surface where there are only u- and d-quarks, which have longer Compton wavelengths than s-quarks.  The latter must be absent from the SQN surface. 
This means that, for $m_s$ in the region of 150 MeV, the charge of the lattice of $3A$ quarks is about $0.3A^{2/3}$.We also computed the time to reach equilibrium in the charge ($Z_N$) from $\tau=Z_N/\dot{Z}$.  The results are collected into Table III.

\vspace*{.25in}
\begin{center}{\bfseries\large{RESULTS}}\end{center}
 
\begin{figure}
\includegraphics[width=3in]{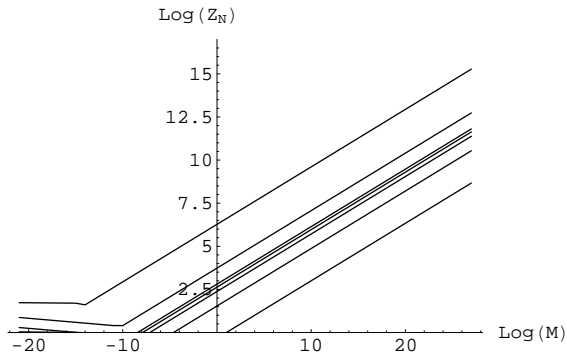}%
\caption{SQN charge $Z_N(M)$. From the top, the 7 curves are in the order described in the text and used for the 7 rows of Table II.   Mass, M, is in grams.}.
 \end{figure}
Figure 1 shows the results as a function of SQN mass (M). We get these results by 

\begin{table}
\noindent
\caption{Parameters for the seven curves of Fig. 1. E is in eV, $\bar{n}_e$ has units of cm$^{-3}$, $\bar{v}_e$ has units of cm/sec. F(E) is  for the non-thermal distribution cases. $F(E)=AE^{-3.05} \gamma/($cm$^2$\,s\,eV) where average temperatures are not known, we assume 100\,K.}

\begin{ruledtabular}
\begin{tabular}{lllr}

 {\bfseries  SQN Location}& {\bfseries   Radiation} &  $\bar{n}_e$&$\bar{v}_e/10^5$ \\

\hline
Solar X-ray  &&&\\ 
 \ \ Flare at 1\,AU&$T=10^3$eV&$7 $&$500$\\ 
 IGM&F(E); Log(A)=7.1 &$10^{-7}$&$60 $\\
 Near Quiet Sun  &$T=0.5$ eV&$7$&$500$\\
IGM\,Pre\,Recombo &CMB $T=0.26$ eV&$287 $&$300 $\\
Galaxy Center &F(E); Log(A)=11.57  &40&60\\
Near Sun(DRB) & F(E); Log(A)=7.64&7&$500$\\
Today (CMB) & $T=2.75$K&$10^{-7}$&$60$\\

\end{tabular}
\end{ruledtabular}

\end{table}
\noindent setting  the rate $(\dot{Z_+})$ at which the SQN loses electrons due to the radiation  equal to the rate $(\dot{Z_-})$ at which it gains electrons from capture of ambient electrons. This yields the equilibrium value of the number liberated ($Z_N$).  We did this for SQN masses ranging from $10^{-20}g$ to $10^{+30}$g.  The bends in the curves of Fig. 1 occur where the Bohr orbit of the most loosely bound electron becomes smaller than the radius of the SQN.  This, as a function of M, occurs at roughly a nanogram.

Our expression for the rate of ionization is

\begin{equation}\label{zplus}
	\dot{Z_+}=4\pi b^2cF_{\gamma}(E'>E)
\end{equation}

 \noindent where $F(E'>E)$ means all the photons with energy $E'$ greater than $E$.  The rate of ionization, $\dot{Z_+}$, as a function of E, decreases with E (as E increases there are fewer photons with greater energy).  Binding energy magnitude, $E_B$, is related to $Z_N$ by $E_B=Z_N\alpha/b$ where b is the maximum of $r_N$ and $a_B/Z_N$ with $a_B$ the hydrogen Bohr radius.  We take two kinds of photon spectra: a power law falloff for the diffuse radiations of settings 2, 5, and 6 and thermal distributions for the other four.  Our parameters for the settings are given in Table II.  Our expression for electron capture is given by

\begin{eqnarray}\label{zminus}
	\dot{Z_-}=4\pi r_N^2 v_en_e(1+f_e)
\end{eqnarray}
\noindent

\noindent where $f_e$ is a focusing factor, Safronov effect [14 ]  taking into account that, with its net positive charge, the SQN attracts electrons so that its cross section for electron capture is larger than geometric.  $f_e$ can be written $E_B/E_e$ with $E_e$ simply the ambient electron's kinetic energy. Other than this effect, our expressions have only geometric cross sections with no reference to the underlying particle physics cross sections.  This is because the latter, off electrons, are on the order of $10^{-25}$ while the surface density of electrons is closer to $10^{-24}$ so that the probability is that, if the incident electron or photon is within the geometric cross section, the reaction will occur.  This approximation would need to be revisited for a setting involving relativistic electrons. These results are discussed in more detail in Abers \textit{et al.} [16].

\begin{table}
\noindent
\caption{\normalsize {Times and Binding Energies. The settings are grouped into three sections: the  three solar system ones together; the two intergalactic ones together; and the universe with just the CMB -- before recombination and today.}} 
\begin{ruledtabular}
\begin{tabular}{lllrr}
{\bfseries Setting}&&{\bfseries M }$^{1/3}\tau_{Eq}(y) $& {\bfseries E}$_B(eV)$ & {\bfseries E}$_B(eV)$ \\
&&&$M>10^{-10}$g&$10^{-21}$g\\
\hline
Galactic Center&&$38 \times10^{-4} $&$4.5$&$52$\\
\,IGM Today: DBR&&$151$&$100$&$760$\\
\hline
Solar system:&&&&\\
\,\,during X-ray flare &&$1.8\times10^{-5}$&$3.5\times10^4$&$3.9\times10^4$\\
\,\,from GDR &&$1.0\times10^{-5}$&$ 0.65$&$12$\\
\,\,Quiet Sun& &$1.7\times10^{-5}$&$12$&$18$\\
\hline
Recombo with CMB &&$2.6\times10^{-7}$&$8.0$&$10$\\
\,Today from CMB& &$92$&$8.7\times10^{-3}$&$0.012$\\

\end{tabular}
\end{ruledtabular}
\end{table}

\vspace*{.25in}
\begin{center}{\bfseries\large{ DISCUSSION
}}\end{center}

Given the results of the previous section, the next question is how can the SQN charge in these seven settings be exploited to find the SQNs.  We give a few directions that might be worth examining.

1.  Direct detection.  Direct detection needs a system that can identify a charge to mass ratio as predicted by the results in the Figure.  We begin with the expected frequency, $dN_{ev}/dt$ of events (passages into or through a detector of area A),   

\begin{equation}
dN_{eV}/dt=n_{SQN}v_{SQN}A
\end{equation}
	
\noindent where n and v are the number density and speed of the SQNs.  Putting in the local DM density (0.3 nucleon masses/cm$^{-3}=n$ divided by M for $n_{SQN}$.) and the galactic virial velocity, gives us that, if we want to instrument an area A and ask how many events we should get in a time $\tau$, the result is $10^{-17}A\tau/M$.  If a square kilometer could be instrumented, nanogram SQNs could be detected at rates up to 100 each second, and one gram ones at about one each year,	

2.  Unique line.  If an SQN were located close enough to a source of high energy gammas whose  energy is above twice the electron mass (i.e., over 1.1 MeV), the SQN would likely emit gammas of characteristic energy about 1.02 MeV.  Madsen has studied vacuum breakdown into electron-positron pairs in such a case [15].  The magnitude of the binding energy of the least bound electron would be $2m_e$, if one waited long enough,  from the balance between energetically favored decay of the vacuum into an electron positron pair with the electron being captured and further ionization by photons,  The time for such vacuum decay increases exponentially as the two thirds power of the SQN mass.  For a given mass, the time decreases as $Z_N^2$.  Thus there will be a range of masses for which the system will radiate gammas of approximately 1.02 MeV.  For this range of masses, the SQN will ionize until the quark-electron total charge is large enough to make the most loosely bound electron bound by 1.02 MeV or more, and large enough to keep the time for vacuum decay of the same order as the time for ionization.  The 1.02 MeV gammas themselves would come from electrons being captured by the SQN at a time when the balance between further ionization by ambient gammas and vacuum production of electron-positron pairs keeps the binding energy of the least bound electrons at or near that value.  Calculations of these rates can be done, if needed, with some precision using work of three of us solving the Dirac equation for the wave functions with binding energies $E_B=-2m_e$, see [17].

3.  SQM in the sun.  If SQM is the the lowest energy state of baryonic matter, a fair amount of it must pass through the sun, but lose energy during the passage leading to capture.  Calculating the 
amount captured needs to take into account the size of the charge that develops on the SQN since that charge, through the Safronov effect, will enhance the rate of SQN energy loss beyond that of the simple geometric cross section usually used in SQN energy loss estimates.  Another new feature in computing the rate of SQN energy loss is that ionizing the SQN passing through the sun will not be governed by electrons being ejected by photons.  The sun contains more electrons than photons, so the SQN charge will be determined by a balance between ionization by fast electrons and deionization by capture of slow electrons.

\vspace*{.25in}
\begin{center}{\bfseries\normalsize{\large{ ACKNOWLEDGEMENTS
}}}\end{center}
VLT very much appreciates a number of very helpful conversations with Demos Kazanas, one with Floyd Stecker on the diffuse gamma ray spectrum and one with M. Alford on effects of Cooper pairing.  Dr. Alford kindly provided a number of insightful comments on the first version of this paper.  DCR and VLT are also grateful to Jonathan Gardner for calling SN-2006gy to our attention, as well as C. Kilbourne for talking WHIM with us.
DAD was supported in part by the U.S. Department of Energy under Grant No. DE-FG03-93ER40757.  WWR was supported in part by the National Science Foundation under Grant PHY-0555544.

\vspace*{.25in}
\begin{center}{\bfseries\large{REFERENCES}}\end{center}

\normalsize{\noindent
\hspace*{.25em} 1. A. R. Bodmer, Phys. Rev. D {\bfseries 4}, 1601 (1971).\\
\hspace*{.25em} 2. E. Witten, Phys. Rev. D {\bfseries 30}, 272 (1984).}\\
\hspace*{.25em} 3. E. Farhi and R. L. Jaffe, Phys. Rev. D {\bfseries 30}, 2379 (1984).\\ 
\hspace*{.25em} 4. A. de Rujula, and S. Glashow, Nature (London) {\bfseries 312}, 734 (1984).\\
\hspace*{.25em} 5. C. Alcock, E. Fahri, and A. Olinto, Astrophys. J. {\bfseries 310}, 261 (1986). \\
\hspace*{.25em} 6.  M. Alford, K. Rajagopal, and F. Wilczek, Phys.
Lett. {\bfseries  B422}, 247 (1998).\\
\hspace*{.25em} 7. P. Jaikumar, S. Reddy ,and A. W. Steiner, Phys. Rev. Lett. {\bfseries 96}, 041101 (2006).\\
\hspace*{.25em} 8. A. L. Watts and S. Reddy, MNRAS L, {\bfseries L63}, 379 (2007).     \\
\hspace*{.25em} 9. J. Sandweiss, J. Phys. G {\bfseries 30}, S51 (2004).   \\
10. P. B. Price,  Phys. Rev. D {\bf 38}, 3813 (1988). \\
11. C. Spiering, \textit{Proceedings of ICRC2001} 1242, (2001). \\
12. W. B. Banerdt \textit{et al.} in \textit{IDM 2004: 5th International Workshop.
on the Identification of Dark}\\
\hspace*{1em
} \textit{ Matter, Edinburgh,}  edited by V. Kudryavtsev (World Scientific, New Jersey, 2005) p. 581. \\ 
13.  E. T. Herrin, D. C. Rosenbaum, and V. L. Teplitz, Phys. Rev. D {\bfseries 73}, 043511 (2006). \\
14. V. Safronov, Ann. Astrophys.{\bfseries 23}, 979 (1960).\\
15. J. Madsen, Phys. Rev. Lett. 100, 151102 (2008).\\
16. E. S. Abers \textit{et al.}, Phys. Rev. D (to be published).\\
17. D. A. Dicus, W. W. Repko, and V. L. Teplitz, Phys. Rev. D{\bfseries 78}, 094006 (2008).

\end{document}